\documentstyle[12pt,aaspp4]{article} 
\begin{document}

\def\htwomin{h_{50}^{-2}}
\def\hmin{h_{50}^{-1}}
\def\Msun{M_\odot}
\def\Lsun{L_\odot}
\def\kms{km s$^{-1}$}
\def\kpc{\hbox{kpc}}
\def\Lya{{\rm Ly}\alpha}
\def\Lyalpha{{\rm Ly}\alpha}
\def\qnot{q_0}
\def\Iab{I_{814,AB}}
\def\Rab{I_{606,AB}}
\def\etal{et al.}
\def\omit#1{}
\def\Deg{${}^\circ$\llap{.}}
\def\Min{${}^{\prime}$\llap{.}}
\def\Sec{${}^{\prime\prime}$\llap{.}}

\title{A Pair of Lensed Galaxies at $z=4.92$ in the Field of CL1358+62
\altaffilmark{1}}

\author{Marijn Franx\altaffilmark{2}}
\author{Garth D. Illingworth\altaffilmark{3}}
\author{Daniel D. Kelson\altaffilmark{3}}
\author{Pieter G. van Dokkum\altaffilmark{2}}
\author{Kim-Vy Tran\altaffilmark{3}}
\author{{\sl To appear in ApJ Letters, 486, L75 (Sept 1997)}}

\altaffiltext{1}{Based on observations taken with the NASA/ESA {\it
Hubble Space Telescope\/} obtained at the Space Telescope Science
Institute, which is operated by AURA under NASA contract NAS5-26555,
and observations obtained at the W. M. Keck Observatory, which is
operated jointly by the University of California and the California
Institute of Technology}
\altaffiltext{2}{Kapteyn Institute, P.O. Box 800, NL-9700 AV, Groningen,
The Netherlands}
\altaffiltext{3}{University of California Observatories / Lick Observatory, 
Board of Studies in Astronomy and Astrophysics, 
University of California, Santa Cruz, CA 95064}

\begin{abstract}
The cluster CL1358+62 displays a prominent red arc in WFPC2 images
obtained with the {\it Hubble Space Telescope}.  
Keck spectra of the arc show $\Lya$ emission
at 7204 \AA, a continuum drop blueward of the line, and several
absorption lines to the red.  We identify the arc as
a gravitationally lensed galaxy at a redshift of $z=4.92$.  It
is the highest redshift object currently known.
A  gravitational lens model was used to reconstruct images of the 
high-redshift galaxy.  The reconstructed image is asymmetric, containing a
bright knot and a patch of extended emission 0\Sec 4 from the knot.
The effective radius of the bright knot is 0\Sec 022 or 130$\hmin$ pc.  The
extended patch is partially resolved into compact regions of star
formation.  The reconstructed galaxy has $I_{AB}$= 24, giving a
bolometric luminosity of $\sim 3\times 10^{11} \Lsun$.  This can be
produced by a star formation rate of 36 $\htwomin$ $\Msun$ yr$^{-1}
(\qnot=0.5)$, or by an instantaneous star burst of $3\times 10^8$
$\Msun$.
The spectral lines show velocity variations on the order of 300 \kms
along the arc.  The Si II line is blue shifted with respect to the
$\Lya$ emission, and the $\Lya$ emission line is asymmetric with a red
tail.  These spectral features are naturally explained by  an outflow model,
in which the blue side of the $\Lya$ line has been absorbed by
outflowing neutral H I.  Evidence from other sources indicates that outflows
are common in starburst galaxies at high and low redshift.
We have discovered a companion  galaxy 
with a radial velocity only 450 \kms different than the arc's.  The
serendipitous discovery of these two galaxies suggests that systematic
searches may uncover galaxies at even higher redshifts.
\end{abstract}


\received{April 11, 1997}
\accepted{June 19, 1997}

\section{Introduction}

The number of spectroscopically confirmed high-redshift galaxies is
increasing rapidly.  Steidel et al.~(1996a,b) demonstrated that
galaxies at redshifts $z\sim 3$ can be found very
efficiently by deep imaging and color selection (the U or B-band
``dropout'' technique).
The absorption to the blue of $\Lya$ by the $\Lya$ forest, and by
source and intergalactic H I beyond the Ly limit, are unique
signatures of high-redshift objects (Madau 1995, McMahon 1997).  Deep
searches for $\Lya$ emitters have been less successful, but have
recently resulted in detections at high redshift (e.g., Hu and McMahon
1996, Djorgovski et al. 1996).  Although most of the photometric
searches have concentrated on galaxies at redshifts $z=3\hbox{-}4$, similar
searches can, in principle, be used to find even higher redshift
galaxies.

Our serendipitous discovery of two galaxies at $z=4.92$ reported here
suggests that $z\ge 5$ galaxies may be found by similar photometric
``dropout'' techniques.  One of the galaxies is strongly
gravitationally lensed and is detected as a very red arc in the
$z=0.33$ cluster CL1358+62.  This cluster was selected for detailed
study of its member galaxies using HST imaging and extensive WHT and
Keck spectroscopy.  An earlier search for arcs in this cluster was
unsuccessful (Le F\`evre et al. 1994), so the presence of the arc in the
HST image was unexpected.  
HST and Keck are able to exploit the magnification provided by the
cluster lens to provide spectroscopic observations of a high redshift
galaxy at unprecedented spatial resolution.

\section{Observations}

We obtained a large mosaic of multicolor WFPC2 images of the cluster
for a detailed study of the cluster members.  The integration time was
3600 s in each of the F606W (``R'') and F814W (``I'') filters.  The
data have been described in Kelson et al. (1997) and van Dokkum et al.
(1997).  A color image of the central part of the cluster is shown in
Figure 1 (Plate 1). 
Several arcs can be seen, of which the long red
arc (B--C) is the brightest, lying $21'' \pm 1''$ from the 
central cD galaxy.  This arc is distorted by the
bright elliptical galaxy north of component B. An additional faint lensed
component D is located still further north.
The arc is strikingly
knotty, with both tangential and radial structure.

The symmetric distribution of the knots suggest that we are seeing a fold
arc. We therefore searched for the expected counterimages. 
We identified a  candidate counterimage  29\Sec 5  east of the cD
galaxy, marked A in Figure 1. 
The colors of the arc and the counterimage are consistent.

Spectra of the arc and the counterimage were obtained
at the W.M. Keck Telescope in 1996 May, June, and 1997 February.
The best spectrum was taken with a long-slit in February, 
using the Low Resolution Imaging Spectrograph (LRIS). 
For this LRIS spectrum, the slit was aligned with the arc (B--C), and
a resolution of 4.3 \AA\ FWHM was attained. The spectrum is shown in
Figure 2 (Plate 2).  The dominant features are the emission line at
7204 \AA\ and the continuum drop to the blue side of this emission
line.  The continuum disappears further in the blue. The spectrum
varies along the arc (e.g., the continuum is enhanced in the bright
knot and the ratio of $\Lya$ to continuum changes).


Figure 3 is a spatially-averaged spectrum showing the emission line
and continuum step at 7204 \AA\ and absorption lines to the red.  We
mark the absorption line features found by Lowenthal et al. (1997) in
an average of 12 high redshift galaxies. Each absorption line can be
identified in the arc spectrum.  These spectral features provide a
secure redshift for the source galaxy ($z=4.92$) and allow us 
to identify the emission feature as $\Lya$.  The redshift of
$z=4.92$ is currently the highest spectroscopically measured for
a galaxy or active nucleus.  The narrow $\Lya$ emission, the absence of
other emission lines, and the broad spatial extent imply that the
continuum is stellar.

The spectrum suggests that the continuum break across $\Lyalpha$ is 
about a factor of 1.9. This is an underestimate, as the
break is very likely diluted by a weak contribution from the bright
cluster galaxy. This galaxy contributes a flux in the I band similar to
that of the arc. The broad band colors, and the spectrum of the bright
knot at the end of the arc suggest that the true flux ratio across the 
break is a factor of 3 or more.

We also obtained spectra of the candidate counterimage in multislit
mode.  It was found to have an emission line at the same redshift as
the arc (Figure 4b), confirming its status as a counterimage of the
arc. In what follows, we refer to the galaxy that is gravitationally
lensed into these arcs as G1.  To our surprise, we found an
unexpected emission line from a faint, red, galaxy (``G2'') that lies
37$''$ W of the arc (Figure 4c).  It was observed by chance through a
slitlet pointed at another galaxy.  It has the same shape and width as
the $\Lya$ emission from the arc. Both $\Lya $ lines have a tail to
the red, and a steep cutoff in the blue.  This signature has been
found previously in two nearby starburst galaxies (Lequeux et al. 1995,
Kunth et al. 1996).  As these authors show, this asymmetry is
caused by an outflow.  Our results indicate that high
resolution spectra can be used to distinguish between $\Lya$ emission
and O [II] 3727 \AA: at high resolution O [II] is resolved into a
symmetric doublet, without the asymmetry displayed by $\Lya$ (Figure
4).

The wavelength peaks of the lines in object G2 and the arcs are very
similar: we measure a wavelength of 7193 \AA\ for G2, versus 7204 \AA\
for the arcs.  Object G2 has a magnitude of $\Iab$= 25.1, and is also
very red (see Figure 1, Plate
1).  Thus the emission is highly likely to be $\Lya$, and G2 is at
$z=4.92$.  After correction for lensing (see below), the
projected distance between G2 and G1 is 200 $\hmin$ kpc, and the
velocity difference is 450 \kms\ (with an unknown error dominated by
the outflow in both galaxies).

\section{Analysis}

The imaging and spectroscopy leave little doubt that we have found a
multiply-imaged, distant galaxy.  We have been able to reproduce the
geometry of the arcs by modeling the cluster as an isothermal
potential.  An additional isothermal potential was added to model the
effect of the elliptical galaxy north of the arc. The general form of
the model is shown in Figure 5 (Plate 3).  This model successfully
reproduces the fold arc and the counterarc. Furthermore, the model
produces only one counterimage, because the source lies outside the
radial caustic.  The Einstein radius of this model is 21$''$.  For an
isothermal model, this implies a velocity dispersion of 970 (990)
\kms\ 
for $q_0$=0.5 (0.05), which is similar to the observed dispersion of
1080 \kms\ for the galaxies (Fisher et al., in preparation).

We use this lens model to reconstruct the intensity distribution of
the source galaxy, G1. The two parts of the fold arc, and the 
counterimage were reconstructed separately. The reconstructions are shown in
Figure 1 with the color information and with higher contrast in
Figure 5 (Plates 1 and 3).  The resolution varies from one
reconstruction to the next, but is particularly high for the west part of
the fold arc where the magnification is large ($5-11 \times$).  The
morphology is a strong function of the resolution.  At the lowest
resolution (with 0\Sec 05 $\times$ 0\Sec 1 pixels in the source
plane), the galaxy appears to consist of a bright knot and a diffuse
extended structure. These features contribute roughly equally to the
total flux.  At the highest resolution, half of the flux of the extended
structure is resolved into small knots.  The bright knot is very
small. It is extended in a PC image, but this is partly due to the PC
point spread function.  After deconvolution and reconstruction, we
find that the half-light radius of the knot is 22 milliarcseconds or
$\approx 130\hmin$ pc ($q_0 = 0.5$). 

Nearby starburst galaxies show comparable irregular morphologies in
the UV (Meurer et al. 1995) with dense knots of star formation.  Meurer
et al.~ refer to these knots as ``super-star clusters'' and suggest
that they are possibly the progenitors of globular clusters.  The
dense knots can contribute between 20 to 50\% of the UV flux.  The
bright knot in G1 is much more luminous than those in nearby starburst
systems, but otherwise the similarity is striking.

\subsection{Star Formation at $z=5$}

The lens model implies that the unlensed magnitudes
of G1 and G2 are  $\Iab$ $\approx$ 24 and 25.4, respectively.  The
corresponding star formation rates (SFR) are 36 and 9 $\Msun$/yr
($q_0$=0.5).  This is based on model predictions by Bruzual and
Charlot (1993) for a stellar population with continuous star formation
at an age of $10^8$ years, and a Salpeter IMF.  An instantaneous burst
model requires a minimum mass of $3 \times 10^8$ $\Msun$.  If dust is
present, the above SFRs would increase, possibly substantially (Meurer
et al. 1997).  Even so, the inferred star formation rate is very high.
Over a period of $10^8$ years, which is only 1/6 of the Hubble time at
$z=4.92$, 20-40\% of the mass of the Galactic bulge can be formed.

Meurer et al. (1995, 1997) suggest that there is a natural upper limit
to the surface brightness of star bursts.  We find that the mean
surface brightness of G1 is low at $1.4 \times 10^{10}$ $\Lsun
\kpc^{-2}$.  However, the surface brightness of the knot in G1 is much
higher, at $3 \times 10^{12}$ $\Lsun \kpc^{-2}$. This is significantly
higher than the 90th percentile upper limit of $2 \times 10^{11}$
$\Lsun \kpc^{-2}$ derived by Meurer et al. (1997), although it is close
to the highest value in their sample.  The bright knot in G1 may appear
extreme because it is unique amongst high redshift galaxies in being
mapped with very high spatial resolution (from the magnification
provided by the cluster lens).  At the usual WFPC2 resolution, the
knot in G1 is undersampled.  This undersampling may pose a problem
for size estimates based on WFPC2 observations of unlensed high
redshift galaxies.

The bright knot in G1 is too massive to evolve into a typical globular
cluster.  If the knot remains bound, it is more likely to end up in
the nucleus of the galaxy, and it is possibly one of the progenitors
of the bulge.  A dynamical mass estimate of the knot would be
valuable.

\subsection{Velocities and Spatial Structure in the Spectrum}

The arc is the first high redshift galaxy for which it is possible to
obtain spatially-resolved spectral information (Figure 6).  The peak
of the $\Lya$ emission varies by about 7 \AA.  This variation is real,
and is larger than expected from positional variations of the knots on
the slit.  The Si II 1260 \AA\ is blue shifted with respect to the
$\Lya$ peak by 300 \kms, and the velocity variation is on the order of
300 \kms. The width of the Si II line is approximately constant at 12
\AA\ (or 2 \AA\ in the rest frame), which is equivalent to a
velocity dispersion of 200 \kms.  However, these velocities should
{\it not} be interpreted as indicative of the gravitational potential
of G1.

The blue shifted Si II absorption, and the asymmetric $\Lya$ emission
are naturally explained by an outflow model. 
Outflowing neutral and ionized material accounts for the Si II
absorption, and absorbs the blue part of the $\Lya$ emission line. The
low continuum blueward of $\Lya$ (Fig. 2) is consistent with the
presence of a significant, blueshifted H I column.  G1 is thus similar
to Mrk 33, a nearby starburst galaxy.  Lequeux et al. (1995) observed
the same blueshifted interstellar absorption lines and asymmetric
$\Lya$ profile in Mrk 33. They verified that the interstellar
absorption lines were blueshifted with respect to the bulk of the
galaxy. Another such case was presented by Kunth et al. (1996).
Furthermore, Heckman and Leitherer (1997) found unambiguous evidence
that UV absorption lines in NGC 1705 are produced by outflowing
material.  In general, outflows are present in most luminous
starbursts at low redshifts (e.g., Heckman, Armus, and Miley 1990).

Outflows may also be the norm for high redshift galaxies: G2 has the same
asymmetric $\Lya$ profile, and the $\Lya$ emitter found by Djorgovski
et al. (1996) has a red tail.  Furthermore, the average spectrum of 12
high redshift galaxies obtained by Lowenthal et al. (1997) shows a
velocity offset between the $\Lya$ emission, and the Si II absorption.
Trager et al. (1997) found a similar blueshift for two galaxies at
$z=4$.

The $\Lyalpha$ emission may also be redshifted with respect to the 
system velocity because of absorption by the $\Lyalpha$ forest. This does
not appear to be the dominant effect for the arc, as the high resolution
spectrum in Figure 4a shows a very strong drop in the blue. Furthermore,
the blue edge of the line shows velocity variations across the slit.
The case is less clear for other high redshift galaxies where the spectra 
are typically at lower resolution.

The apparent ubiquity of outflows in galaxies with substantial SFRs,
casts doubt on the suggestion by Steidel et al. (1996a) that the
regular rotation caused the
broadening of the interstellar absorption lines in high redshift
galaxies.
This is unfortunate
since it would be very useful to find a method of measuring the mass of
high redshift galaxies.  
The large width (200 \kms) of the  Si II line towards the
bright knot in the galaxy is another argument against a rotational
origin for the observed motions.  It is hard to reconcile a systematic
velocity variation of this magnitude across the knot with the small
size of the knot.  
The velocity variations in the $\Lya$ and Si II
across the arc are probably due to differences in the
outflow velocity and absorption column density.  
The bottom line is that wavelength gradients
in $\Lya$ or the interstellar absorption lines cannot be used
to infer reliably the mass of such star forming galaxies.

\section{Conclusions}

We have discovered a red arc and a counterimage of a distant galaxy
that is gravitationally lensed by the cluster CL1358+62 at $z=0.33$.
The lensed galaxy and a second companion galaxy have secure redshifts
of $z=4.92$.  The arc has an irregular structure, and image
reconstructions show that most of the flux comes from very small
knots.  The morphology of the reconstructed galaxy is very similar to
that of nearby starburst galaxies in the UV (Meurer et al. 1995), which
are also very irregular with dense star-forming knots. The surface
brightness of the brightest knot in the lensed galaxy is higher than
the 90th percentile upper limit derived by Meurer et al. (1997) for
nearby and distant starbursts.  The mass and size of this bright knot
suggest that it might be a ``bulge-building'' subclump.  It remains
to be seen whether this lensed galaxy (G1) is typical of the population of
starbursting galaxies at $z\sim5$.

The irregular structure of G1 can be contrasted with the $z=3$
galaxies studied by Giavalisco et al. (1996). These authors found that
many of those galaxies are compact with little substructure.  The high redshift
galaxies in the Hubble Deep Field are more similar to our
reconstructed images of G1
(e.g., Lowenthal et al. 1997, Steidel et al. 1996b).  The morphology of
G1 is a strong function of resolution: at the lowest resolution, half
of the flux appears to originate from a smooth, extended
component. Half the flux of this component is resolved into knots at
the highest resolution.  Other high redshift galaxies might
display similar irregular morphologies if they were to be imaged at much
higher resolution.

Our results, and the data presented by others, indicate that radial
outflows dominate the kinematics of the absorption line gas in high
redshift galaxies.  These radial outflows are a general feature of
luminous starburst galaxies in the nearby universe.  Radial outflows
of hundreds of \kms \ can quickly disperse the reservoir of gas from
which the stars formed.  The rest-UV appearance of these galaxies is
likely to be dominated by short-lived areas of intense star
formation. IR observations are needed to characterize the overall
morphology, since the UV structure will not be representative of the
distribution of older stars in these galaxies.

Our serendipitous discovery of the two galaxies at $z=4.92$ raises the
question of the surface density of such objects. Their unlensed $\Iab$
magnitudes are estimated to be 24 and 25.4. The results by
Lanzetta et al. (1996) suggest an upper limit of $z>4$ galaxies of 0.4
arcmin$^{-2}$ for $\Iab < 26$.
Madau et al. (1996) found $B$ drop-out candidates in the Hubble Deep Field with
$\Iab$ magnitudes of 26 or higher. 
The $\Rab-\Iab$ color of the arc is in good agreement
with the models of high redshift starbursts (Madau et al. 1996), 
which indicates that 
candidates might be found by shifting the drop-out techniques further
into the red.
Finding such objects with emission line searches may be more difficult.
While the unlensed $\Lyalpha$ flux of the arc is comparable to the fluxes 
for the $\Lyalpha$ selected galaxies found by Hu and McMahon (1996), it is
comparable to the rms noise in the deep field searches for 
$\Lyalpha$ (e.g., Thompson, Djorgovski, and Trauger, 1995).

\acknowledgements

We thank Dan Fabricant and Tim Heckman for their comments on the manuscript.
We appreciate the valuable comments of an anonymous referee.
The assistance of those at STScI who helped with the acquisition of the 
HST data is gratefully acknowledged. We also appreciate the effort of 
those who developed the Keck telescopes and the LRIS spectrograph,
and those who supported our efforts at Keck.  Support from STScI grants
GO05989.01-94A, GO05991.01-94A, and AR05798.01-94A is gratefully acknowledged.

\newpage
\onecolumn

{\parindent=0pt
Figures 1,2,5 in jpeg format. For high quality images see \hfill\break
http://www.astro.rug.nl/~franx/papers/arc 
or ftp://www.astro.rug.nl/preprint/238.ps.gz
}
\vskip 0.5truecm
\figcaption{[Plate 1]
\label{wfpc}
HST image of the center of the cluster CL1358+62 at $z=0.33$.  The
cluster was imaged in the F606W and F814W passbands. The red arc is
conspicuous, and is marked B and C.  Other images of the lensed galaxy
at $z=4.92$ are marked (A and D).  The reconstructions of the images
A-C are displayed at the bottom at a highly expanded scale. 
These images may be compared with the lens geometry and lens models shown in
Fig. 4 (Plate 3).
The bright knot is unresolved in the WF reconstruction of C, and only
marginally resolved in the reconstructed PC image shown to the
right. The inset at the upper right displays at enhanced contrast
a second  galaxy, G2, that was found serendipitously 37\arcsec \ west of the
arc. 
}
\figcaption{[Plate 2]
\label{spectwod}
The best Keck spectrum of the red arc is displayed at the top. 
The spectrum is dominated by the emission line at $\lambda \approx
7200$ \AA, which we identify as $\Lya$. The bright knot in the arc
is visible in the lower part of the arc spectrum. The distance from
the brightest knot is indicated along the vertical axis.
The Si II absorption line at 1260 \AA\ can be identified at the red end
of the spectrum.
The two bottom panels show the expanded region around $\Lya$ and Si
II. Sky lines redward of $\Lya$ introduce additional noise in the
spectra, and are indicated by $\earth$. The $\Lya$ emission varies
significantly along the arc. The equivalent width of $\Lya$ is 21
\AA\ near the middle of the arc, but is low at 7 \AA\  at the bright
knot. The continuum of the knot is strongly absorbed blueward of
$\Lya$, but less so for the arc as a whole.
Due to the strong curvature of the arc, the bright knot at the east
side of the arc was not included in the spectrum.
}

\plotone{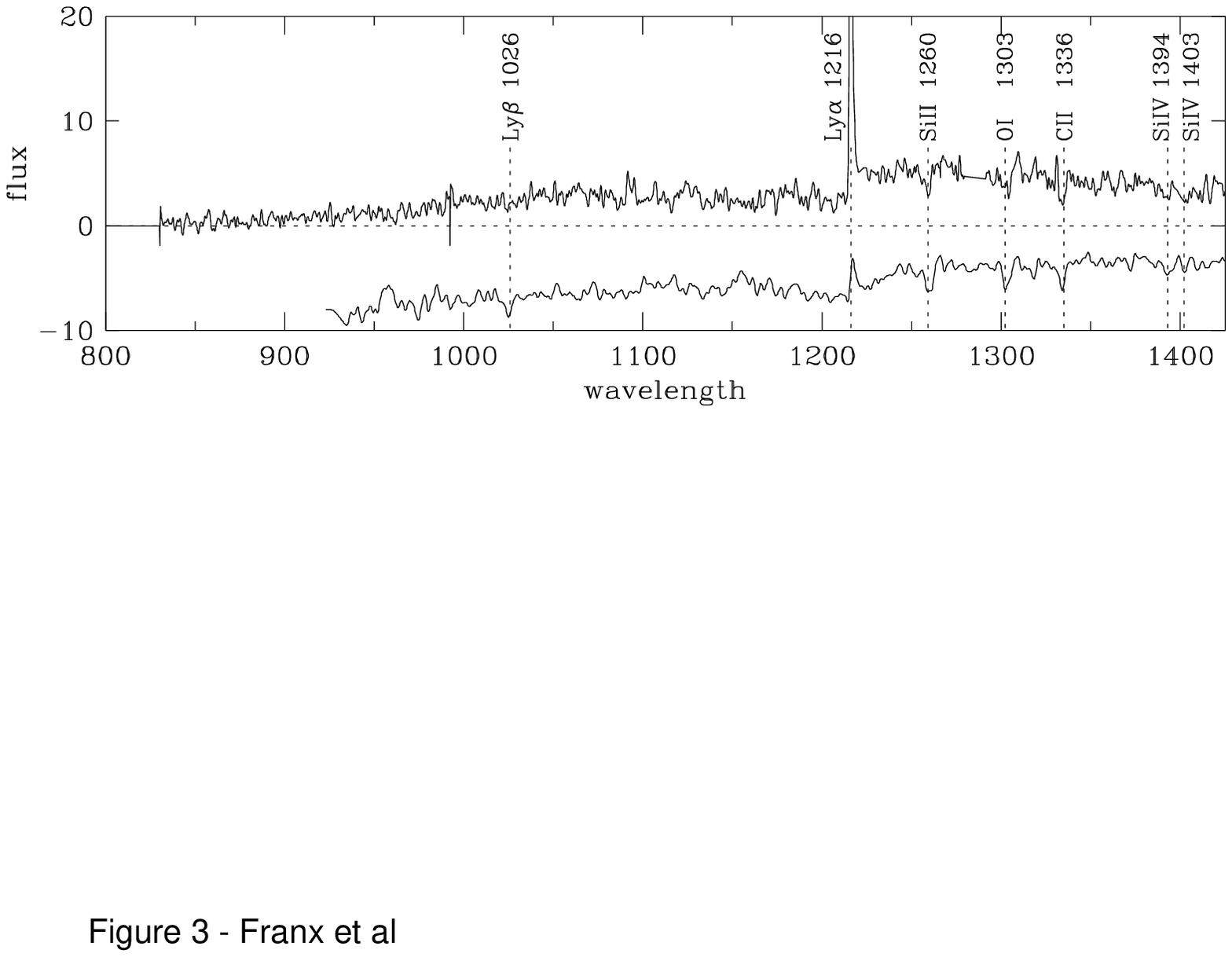}
\figcaption[fig3.ps]{
\label{speconed}
The averaged spectrum of the arc. 
The $\Lya$ emission line, and the discontinuity across the emission
line are clearly visible.
For comparison, the average spectrum of 12
galaxies at $z=3$ (Lowenthal et al. 1997) is shown at the bottom.
 All conspicuous interstellar
absorption features in this average spectrum are marked, and can be
identified  in the spectrum of the arc. These confirm that the arc has a
redshift of 4.92. The continuum has a weak contribution from the
nearby elliptical galaxy, which dilutes the lines somewhat.
}

\plotone{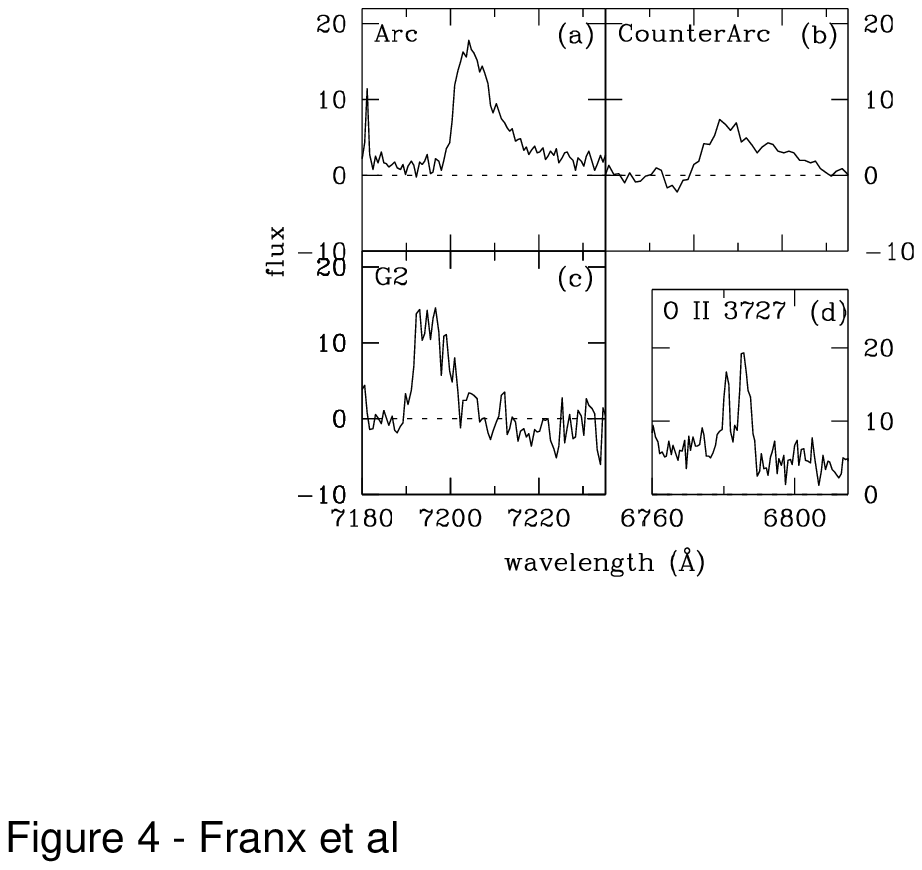}
\figcaption[fig4.ps]{
\label{specslyalpha}
The $\Lya$ emission lines for the arc, the counterarc, and galaxy G2.
The arc and the counterarc have the same redshift.
The emission line of G2 has a similar red tail and width, strongly
suggesting that the emission line is also $\Lya$.
Panel d) shows O [II] 3727 \AA\ emission from a field galaxy at the same
spectral resolution.
The 3727 \AA\ emission is clearly resolved into a doublet, 
and cannot be mistaken for $\Lya$.
}

\figcaption[figure5.ps]{[Plate 3]
\label{recon}
The lens model is shown at the top. The multiple images of the lensed
galaxy are indicated. The galaxy was assigned
three different levels of intensity. These are mapped into the image plane.
The black curves indicate the critical lines in the image plane. The
light curves indicate the caustics in the source plane. The source
is indicated by G1.
The middle row  shows the reconstructions of images A-C at high
contrast.
The resolution varies greatly, and is best in the direction of
greatest magnification.
 It is generally the highest in reconstruction C.
The differences in resolution cause the apparent difference in
morphology.
Image A appears to consist of a bright knot, and extended emission to
the south. This extended emission is resolved into knots in image C.
The brightest of these knots is also visible in image B.
The bottom row shows the same reconstructions after smoothing with the
typical  WFPC2 resolution. Most of the differences
between the reconstructions have disappeared.
}

\plotone{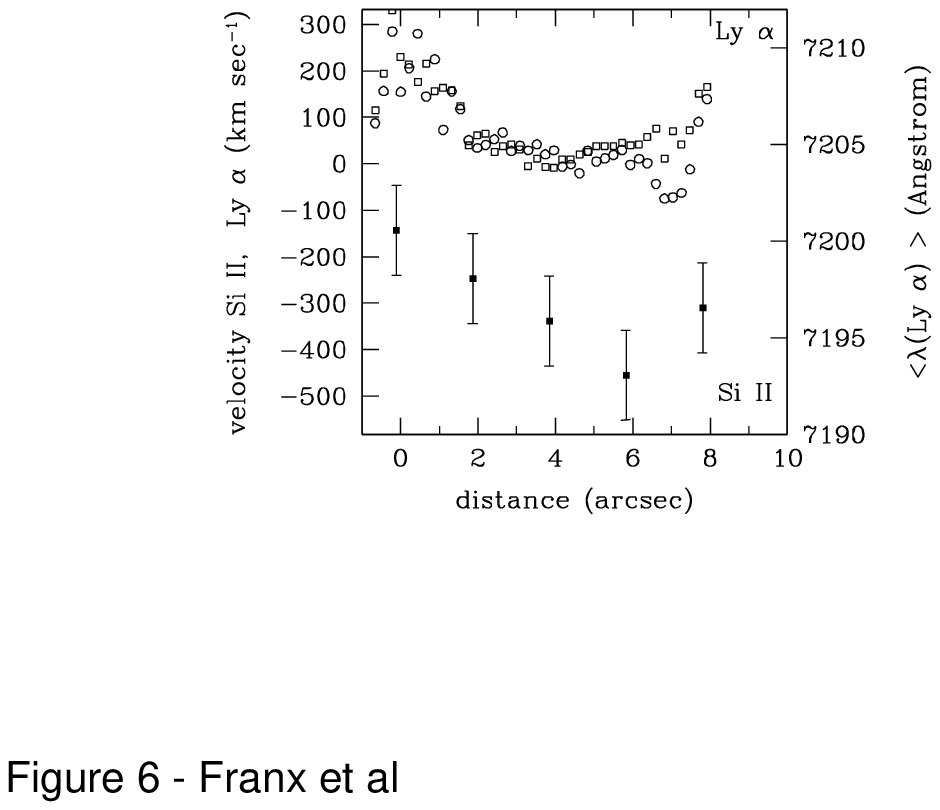}
\figcaption[fig6.ps]{The velocity variations across the arc in the
$\Lya$ emission line and Si II 1260 \AA\ absorption line. 
The open squares and circles indicate $\Lya$ taken with two long slit
exposures.
The closed squares are Si II.
The features show velocity variations on the order of 300 \kms. The Si II
line is systematically blueshifted with regards to the $\Lya$ emission.
The gas kinematics are most likely dominated by an outflow,
which absorbs the blue $\Lya$ emission and causes the
blue shifted Si II absorption.
}

\end{document}